\documentclass{aa}  
\usepackage{graphicx, natbib}
\usepackage{txfonts}
\usepackage{hyperref}
\usepackage{rotating}

\begin{document}

   \title{First determination of $s$-process element abundances in pre-main sequence clusters}

   \subtitle{Y, Zr, La, and Ce in IC 2391, the Argus association, and IC 2602\thanks{Based on observations under ESO programs 080.D-2002 and 082.C-0218}}

   \author{V. D'Orazi
          \inst{1,2, 3}
          \and
          G.M. De Silva\inst{4,5}     
          \and
          C.F.H.  Melo\inst{6}
                 }

   \institute{INAF Osservatorio Astronomico di Padova, vicolo dell'Osservatorio 5, I-35122, Padova, Italy\\
              \email{valentina.dorazi@oapd.inaf.it}
         \and
        {Department of Physics and Astronomy, Macquarie University, Balaclava Rd, NSW 2109, Australia}
        \and
        {Monash Centre for Astrophysics, School of Physics and Astronomy, Monash University, Clayton, VIC 3800, Australia}
         \and
        {Sydney Institute for Astronomy, School of Physics, The University of Sydney, NSW, 2006 Australia}
         \and
         {Australian Astronomical Observatory, 105 Delhi Rd, NSW 2113, Australia}
         \and
        {European Southern Observatory, Casilla 19001, Santiago 19, Chile}
         }

   \date{Received ; accepted }

 
  \abstract
   {Several high-resolution spectroscopic studies have provided compelling observational evidence that open clusters display a decreasing trend of their barium abundances as a function of the cluster's age. Young clusters (ages $\lesssim$ 200 Myr) exhibit significant enhancement in the [Ba/Fe] ratios, at variance with solar-age clusters where the Ba content has been found to be [Ba/Fe]$\sim$ 0 dex. Different viable solutions have been suggested in the literature; nevertheless, a conclusive interpretation of such a peculiar trend has not been found. Interestingly, it is debated whether the other species produced with Ba via $s$-process reactions  follow the same trend with age.}
   {Pre-main sequence clusters ($\approx$ 10-50 Myr) show the most extreme behaviour in this respect: their [Ba/Fe] ratios can reach 0.65 dex, which is higher than the solar value by a factor of four. Crucially, there are no investigations of the other $s$-process species for these young stellar populations. In this paper we present  the first determination of Y, Zr, La, and Ce in clusters IC 2391, IC 2602, and the Argus association. The main objective of our work is to ascertain whether these elements reveal the same enhancement as Ba.}
   {We have exploited high-resolution, high signal-to-noise spectra in order to derive abundances for Y, Zr, La, and Ce via spectral synthesis calculations. Our sample includes only stars with very similar atmospheric parameters so that internal errors due to star-to-star inhomogeneity are negligible. The chemical analysis was carried out in a strictly differential way, as done in all our previous investigations, to minimise the impact of systematic uncertainties.}
   {Our results indicate that, at variance with Ba, all the other $s$-process species exhibit a solar scaled pattern; 
these clusters confirm a similar trend discovered in the slightly older local associations  (e.g. AB Doradus, Carina-Near), where only Ba exhibit enhanced
value with all other $s$-process species being solar. We have discussed  several possible explanations such as  chromospheric effects, departures
from the LTE approximation, or the activation of a different
nucleosynthesis chain. We cannot currently provide the definite answer to this question and future investigations from theoretical and observational perspectives are sorely needed.}
   {}


\maketitle
%

\section{Introduction}
Open clusters (OCs) offer the best examples in nature of a simple stellar population, which is 
a stellar aggregate comprised of coeval stars with the same initial chemical composition. For this reason, they have been extensively exploited to address several fundamental topics over the years from stellar evolution and nucleosynthesis models, to Galactic chemical evolution, up to the search for extrasolar planetary systems, to name a few.
They are indeed very well-known objects. Open clusters do not exhibit the age-metallicity relation and display a radial metallicity gradient (see e.g. \citealt{netopil16}): the presence of old, super-solar metallicity clusters (e.g. NGC 188 and NGC 6791; \citealt{friel10}; \citealt{bragaglia14}) and young subsolar metallicity clusters (the nearby star forming regions; \citealt{santos08}; \citealt{dr09}; \citealt{spina14}) suggests that the actual site of formation is more important than when the OC formed in terms of disc chemical evolution imprinted on these stars. Other elements ($\alpha$, p-capture, and Fe-peak elements) follow iron trends very well, with OCs having [X/Fe] ratios independent of cluster age and galactocentric radius.
Moreover, at variance with their older siblings, the globular clusters, they are very homogeneous and do not exhibit internal variations in light elements (e.g. \citealt{desilva06}).
However, OCs are very peculiar in terms of the heavy-element abundances, in particular regarding those elements produced in the slow ($s$) process of neutron capture 
reactions\footnote{Elements heavier than iron are mostly produced via neutron capture reactions. Two distinct processes are responsible for these heavier nuclei: the slow ($s$) neutron capture process during stellar He burning and the rapid ($r$) neutron capture process (\citealt{burbidge57}; \citealt{cameron57}; \citealt{kappeler11}; \citealt{thielemann11}). AGB stars are the main producers of $s$-process elements, while $r$-process is probably related to supernova explosions. Slow and rapid are defined with respect to the $\beta$-decay timescale.}. 

In recent years, a growing body of evidence has shown that there is a clear trend of the barium (Ba) abundance with the cluster's age.
\cite{dorazi09} first reported the discovery of a decreasing trend for [Ba/Fe] ratios with increasing  open cluster age (both for dwarfs and giants): relatively old clusters (e.g. M67, age $\approx$ 3.5 - 4.5 Gyr; \citealt{sarajedini09}) exhibit solar Ba content; younger clusters (age $\lesssim$ 200 Myr) can reach levels of Ba enhancements up to $\sim$ 0.6 dex. It is basically impossible  to reconcile  this huge value with any standard nucleosynthesis and galactic evolution models, 
which has to account for an incredible amount of Ba production in the last 100 Myr or so. After the pioneering work by \cite{dorazi09}, this very peculiar behaviour has been since confirmed by other authors (e.g. \citealt{jacobson11}; \citealt{mishenina13}). 
Interestingly enough, the question of whether all the other $s$-process elements  follow Ba is still controversial. We  expect that especially those species -- in addition to Ba -- that comprise  the second peak of the $s$-process path, namely La and Ce, should  in principle share the same pattern. We recall that Ba, La, and Ce are defined as heavy s-process elements (hs), whereas the first peak $s$-process elements (Y, Zr) are defined as light $s$-process elements (ls).
\cite{maiorca11}, by analysing 19 OCs with an almost complete overlap with the sample by \cite{dorazi09}, claimed that Y, Zr, La, and Ce also exhibit the rising trend 
for young clusters, though to a  lesser degree  than Ba.  However, \cite{dorazi12} did not confirm these findings, noting that in three young associations (AB Dor, Carina-Near, and Ursa Major) only Ba is enhanced, whereas all other $s$-process species display a solar scaled pattern. Similar results, suggesting a lack of significant trends with age for Y, Zr, and La, were obtained by \cite{jacobson13} based on a sample of 50 OCs, spanning an age range 
between $\approx$ 700 Myr and 10 Gyr.
\cite{yong12}  confirmed the age trend for Ba, though as not as strong as \cite{dorazi09}, and claim to have detected a mild trend for Zr abundances; crucially, they have reported an opposite trend for La, with older clusters showing larger La contents (but see discussion in \citealt{jacobson13} and in Section~\ref{sec:results} of this manuscript).
More recently, \cite{reddy15} have published a comprehensive study for n-capture species in a sample of local associations; they found that all the heavy elements (Y, Zr, La, Ce, Nd, Sm, and Eu) exhibit solar ratios, i.e. [X/Fe] $\approx$ 0, while Ba is overabundant by about 0.2$-$0.3 dex. This is a corroboration of previous results by \cite{dorazi12}.
Recently, \cite{mishenina15} confirmed the trend of Ba overabundance in OCs and suggested that the intermediate ($i$) neutron-capture process might be a viable explanation. 
The $i$-process, which is characterised by neutron densities of $\approx$10$^{15}$ neutrons cm$^{-3}$, has been introduced by \cite{cowan77} and it seems to be triggered by the ingestion of H in the He-burning layers. The first observational evidence was found by \cite{herwig11}
in the post-AGB Sakurai's object, but the stellar source for the activation of the $i$-process is still matter of debate. This mechanism can in principle result in Ba production that is much more efficient than that for other $s$-process species (see \citealt{mishenina15} for details).
It is also noteworthy in this context that the Ba overabundance characterising young populations is not confined to open cluster stars;    isolated young stars in the field also reveal a similar pattern (see e.g. \citealt{desidera11}). 

With the aim to further investigate this debated topic, we present in this paper abundances for Y, La, Zr, and Ce in the so-called pre-main sequence clusters (age roughly $\approx$ 10 $-$ 50 Myr). 
The age range is particularly critical because these are the clusters for which the Ba enhancement is more extreme, reaching values more than a factor of four above the solar content.
The three young clusters IC 2391, Argus, and IC 2602 have reported ages of 53$\pm$10 (\citealt{barrado99}) and 35$\pm$10 (e.g. \citealt{randich01}) for IC2391/Argus and IC 2602, respectively. 
All previous investigations of $s$-process elements other than Ba were focussed on much older open clusters and associations; 
this is the first time that such young clusters have been analysed in terms of their heavy element content. 
The paper is organised as follows. In Section~\ref{sec:obs} we describe details on observations, data reduction, and spectral analysis, while results are presented and discussed in Section~\ref{sec:results}. A summary ends the manuscript (Section~\ref{sec:summary}).


\section{Observations, data reduction, and analysis}\label{sec:obs}

High-resolution and high signal-to-noise (S/N) spectra of members of the Argus association and IC 2391 were observed with 
UVES (\citealt{dekker}) located at UT2 of VLT 
in the framework of program ID 082.C-0218 (PI Melo). We refer the reader to our previous work (\citealt{desilva13}) for details on sample selection, observational strategy, and data reduction procedures. From that original sample we have selected for the current purpose only stars within $\sim$ 200 K of the solar effective temperature. This choice
allows us to minimise the impact of different atmospheric parameters on our internal accuracy. Most crucially, very young stars cooler than $\sim$ 5400 K have been 
found to have departures from LTE for many species, including iron and titanium (e.g. \citealt{dr09}; \citealt{schuler10}), which are key species in the determination of spectroscopic parameters (effective temperature and surface gravity).
For IC 2602, we  have included two solar-type, slow rotating (vsin$i$ $<$ 15 km s$^{-1}$) stars originally observed under program 080.D-2002 (PI Mendez) with the FEROS spectrograph (R=48,000; \citealt{kaufer99}). The spectra provide an almost complete coverage from 3500 \AA ~to 9200 \AA ~with signal-to-noise (S/N) ratios per pixel of 100 at 6000 \AA. All spectra were reduced using the dedicated FEROS data reduction software implemented in the ESO-MIDAS system.

\begin{figure}
\centering
\includegraphics[width=\columnwidth]{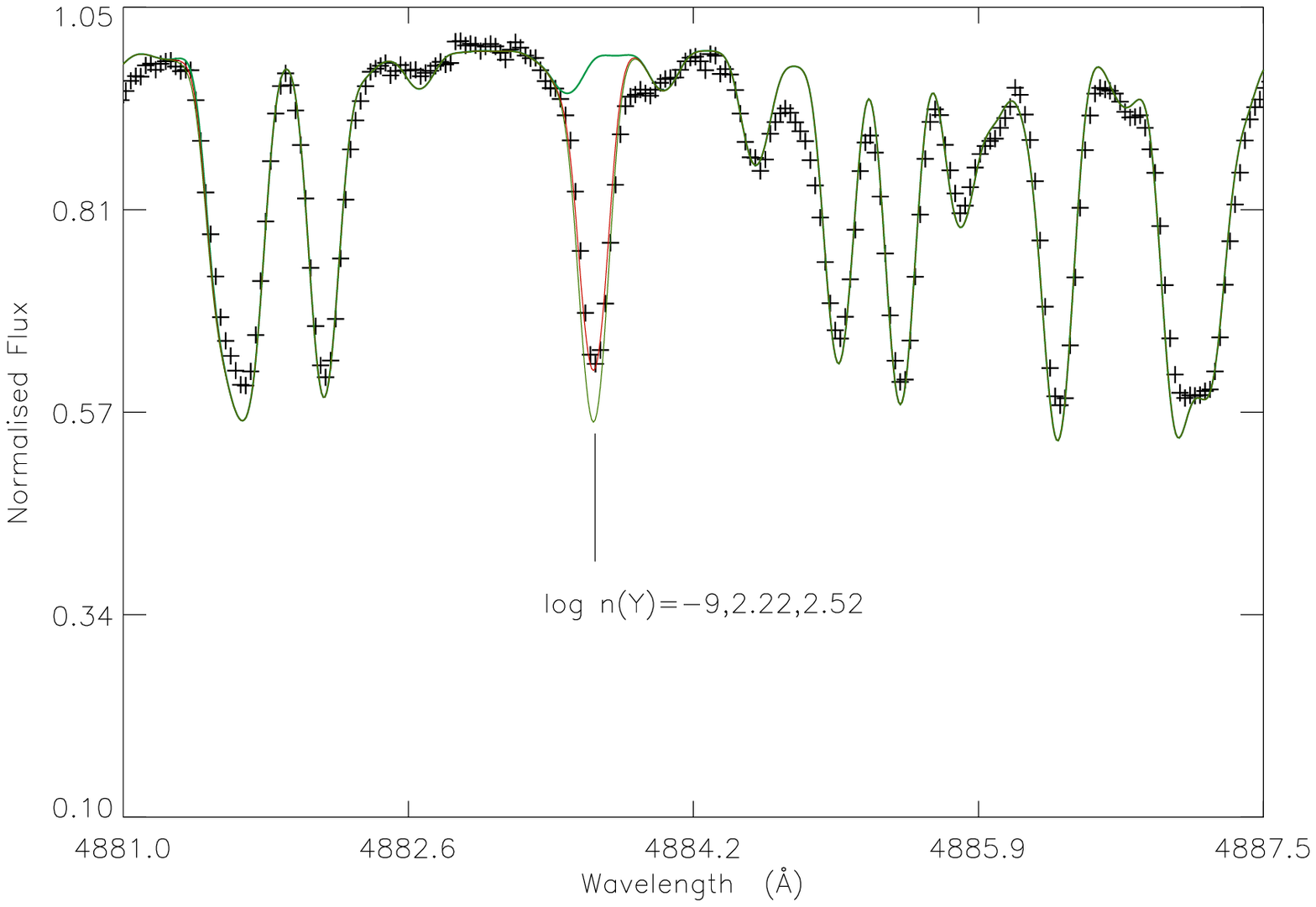}
\includegraphics[width=\columnwidth]{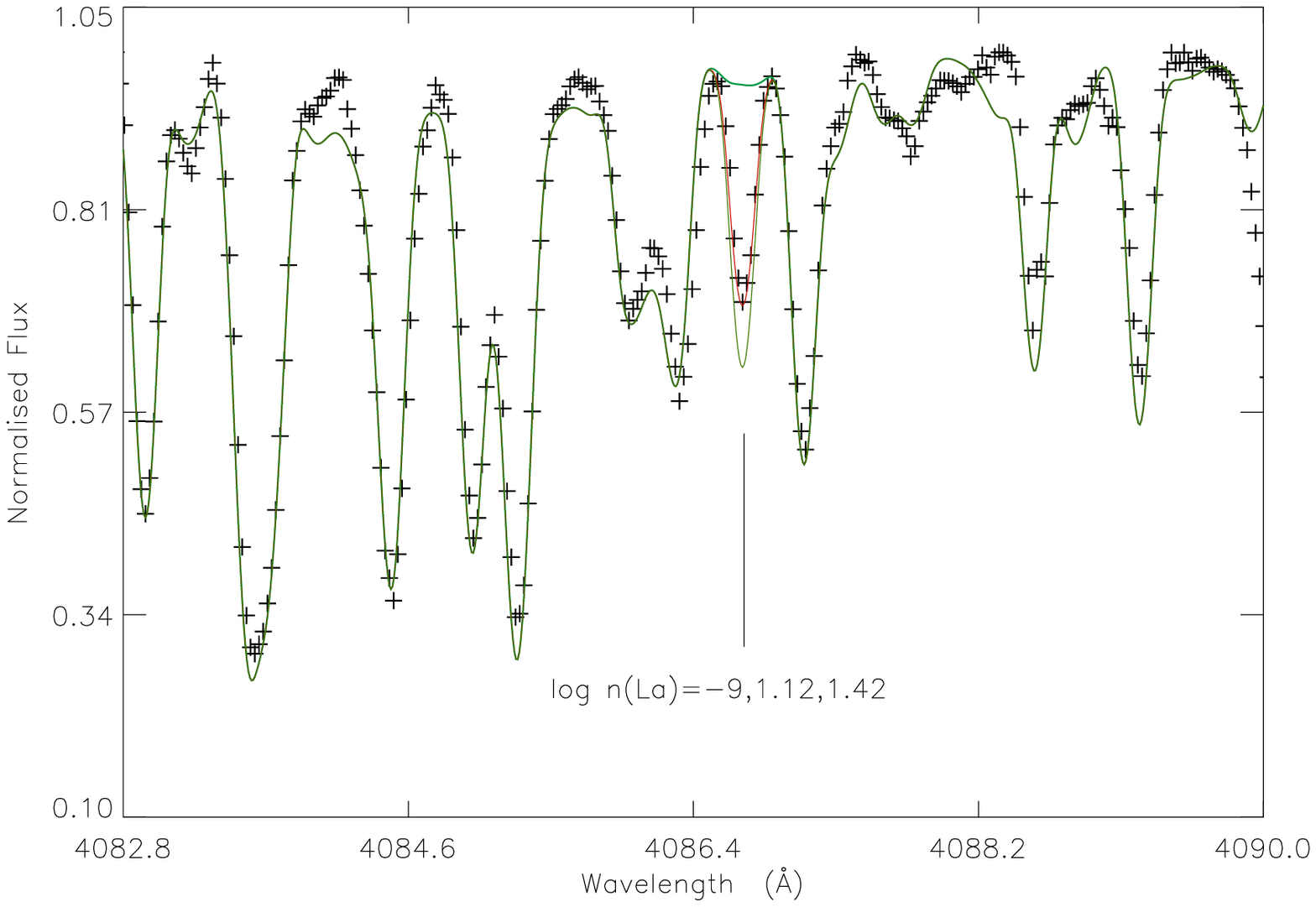}
\caption{Example of spectral syntheses for the Y~{\sc ii} line at 4883 and for the La~{\sc ii} line at 4087 \AA. The star is PMM 1142 belonging to IC 2391.}\label{f:synth1}
\end{figure}

Stellar parameters and metallicity for the Argus association and IC 2391 stars were retrieved from \cite{desilva13}, whereas we have adopted
values published by \cite{dr09} for stars R66 and R70 belonging to IC 2602. 
Since we have already checked that no major systematic uncertainties affect our analysis (which was carried out in a very similar way),
as shown in \cite{desilva13} this choice does not affect our conclusions. There is an excellent agreement for stars in common between the two papers (see \citealt{desilva13} for details).
Abundances for neutron-capture species have been derived via spectral synthesis calculations using the code {\sc MOOG} (\citealt{sneden73}, 2014 version) 
and the Kurucz grid (\citealt{castelli04}) of model atmospheres, as was done in our previous works (\citealt{dorazi12}; \citealt{desilva13}). 
We  carried out a very careful selection of strong, isolated, and unblended spectral lines for the key species.
We have thus exploited the following lines: 4398.01 \AA\  and 4883.68 \AA~for Y~{\sc ii};  4050.32 \AA\  and 4208.98 \AA~for Zr~{\sc ii};  
3988.51 \AA\  and 4086.71 \AA~for La~{\sc ii}; and 3999.24 \AA\  and 4073.74 \AA~ for Ce~{\sc ii}. 
An example of spectral synthesis computation is shown in Figure~\ref{f:synth1}  for Y and La.
Hyperfine structure has been taken into account only for La, following prescriptions by \cite{lawler01}. For the other lines under scrutiny in this work the assumption of a single-line treatment is robust, as previously demonstrated (see also \citealt{dorazi12}). 

Barium abundances have  already been published in our previous works, thus we refer to \cite{desilva13} for IC 2391/Argus, whereas for IC 2602 Ba values are from
\cite{dorazi09}.
Solar abundances are given in Table~\ref{t:sun} and were used to calculate [X/H] ratios throughout this manuscript; the error bar reported in this table is the r.m.s from the average abundance as given by the two spectral features for each species. 
We  exploited a solar spectrum acquired with FEROS to perform the comparison with synthetic spectra in order to gather abundance values, adopting the following parameters: $T_{\rm eff, \odot}$=5770K, log$g_\odot$=4.44 dex, $\xi_\odot$=0.90 km s$^{-1}$, and A(Fe~{\sc  i})$_\odot$=7.50 dex. A Kurucz grid of model atmospheres along with MOOG code have been used for consistency with the chemical analysis of our sample stars.
In Table~\ref{t:sun} we also report the photospheric abundances from \cite{grevesse96} and  \cite{asplund09}, along with meteoritic abundances (\citealt{lodders03}).

\begin{table}
\centering
\renewcommand{\tabcolsep}{0.1cm}
\caption{Solar abundances from the present study compared to abundances by \cite{grevesse96}, \cite{asplund09}, and the meteoritic values
 (\citealt{lodders03})}.\label{t:sun}
\begin{tabular}{lcccr}
\hline\hline
Species & This study & Grevesse+96 & Asplund+09 & Meteoritic \\
\hline
              &                   &                       &                     &                 \\
 Y          & 2.19$\pm$0.03 & 2.24$\pm$0.03 & 2.21$\pm$0.05 & 2.17$\pm$0.04 \\
 Zr         & 2.55$\pm$0.04 & 2.60$\pm$0.03 & 2.58$\pm$0.04 & 2.53$\pm$0.04 \\
 La        &1.05$\pm$0.01 & 1.17$\pm$0.07  & 1.10$\pm$0.04 & 1.17$\pm$0.02 \\
 Ce       & 1.58$\pm$0.01 & 1.58$\pm$0.09 & 1.58$\pm$0.04 & 1.58$\pm$0.02\\
             &                          &                          &                          &                          \\
\hline\hline
\end{tabular}
\end{table}

\subsection*{Error budget}
Two source of uncertainties affect our abundances:  errors related to the best fit procedures, which account for 0.10 - 0.12 dex depending on the spectral features under consideration, and  uncertainties related to the atmospheric parameters ($T_{\rm eff}$, log$g$, $\xi$, and input metallicity [A/H]). 
The impact of the uncertainties on the atomic parameters, i.e. log$gf$, should  instead be almost negligible since our analysis is strictly differential with respect to the Sun and our sample stars are very similar to the Sun in terms of atmospheric properties.
For the second kind of errors we  adopted the values given in \cite{desilva13} and \cite{dr09}, from which we  retrieved the stellar parameters and iron abundances, and checked the corresponding variations on our abundances for Y, Zr, La, and Ce. Because they are independent, we  then added in quadrature both error estimates, i.e. our total internal uncertainties reported in Table~\ref{t:results}.  As  can be seen, random errors are always below 0.20 dex, with the exception of Ba. Because the lines are very strong and close to the saturated part of the curve of growth, the impact of microturbulence velocities is largely responsible for the corresponding uncertainty.
We stress that given the consistency with previous investigations from our group, no major systematic uncertainty affects our analysis and conclusions.


\section{Results and discussion}\label{sec:results}

Final abundances for $s$-process species Y, Zr, La, and Ce are given in Table~\ref{t:results}, along with the corresponding total internal uncertainties (see Section \ref{sec:obs}).
For completeness, we also list the  stellar parameters, metallicity from Fe~{\sc ii} lines (because all the species under investigation are analysed from ionised lines), and Ba abundances from our previous works (\citealt{desilva13}; \citealt{dorazi09}). 
Average abundances for IC 2391 are [Y/H]=0.01$\pm$0.02, [Zr/H]=0.01$\pm$0.01, [La/H]=0.06$\pm$0.01, and [Ce/H]=0.07$\pm$0.01. For the Argus association, 
which is known to comprise also IC 2391, we find the following values: [Y/H]=0.07$\pm$0.07, [Zr/H]=0.01$\pm$0.03, [La/H]=0.14$\pm$0.05, [Ce/H]=0.06$\pm$0.15. 
Finally for IC 2602 we obtained [Y/H]=0.08$\pm$0.04, [Zr/H]=0.06$\pm$0.01, [La/H]=0.09$\pm$0.04, [Ce/H]=0.01$\pm$0.01. 
 Thus our main result is that {light and heavy $s$-process elements exhibit solar abundances}. When considering Fe abundances, as reported for each star 
in Table~\ref{t:results}, we found  solar scaled [X/Fe] ratios once observational uncertainties had been taken into account. These [X/Fe] values are  displayed in Figure~\ref{f:sproc}, where we show abundances for Y, Zr, La, and Ce as a function of the cluster age. The plot includes values from the present work (red triangles for IC2391 and Argus, and blue diamond for IC 2602), along with previous determinations (green asterisks and empty black squares are for data from \citealt{dorazi12} and \citealt{maiorca11}, respectively). The inclusion of our new measurements in this plot confirms that there is no clear evidence for a rising  trend with age for $s$-process elements
(apart from Ba, see discussion below). 

 None of the correlations is statistically significant, as previously found by \cite{jacobson13}. These authors have found that their data show an increasing [Ba/Fe] with decreasing age 
(similarly to \citealt{dorazi09} and \citealt{yong12}); on the other hand, there is no correlation between La abundance and the  cluster's age  (contrary to \citealt{maiorca11} and in agreement with our results).
It is noteworthy in this context that the evidence reported by Maiorca and collaborators that [La/Fe] ratios follow Ba, by displaying a trend with the age, is actually quite weak; 
this conclusion is based on a sample of only six clusters and the linear correlation coefficient $r$=$-$0.29 is not statistically meaningful. 
Interestingly enough, \cite{yong12} has obtained an opposite trend of [La/Fe] as a function of the OC's age; however, as extensively discussed by \cite{jacobson13}, the trend disappears once systematic
uncertainties are taken into account. Thus, they  conclude that none of the available data from the available samples  show evidence for age trends in La abundance (\citealt{jacobson13}).
The same hold for Zr abundances reported from the three studies.
In summary, our findings fit well in a complex picture whereby only Ba abundances show compelling evidence for an increasing trend as  the cluster's age decreases.

\begin{figure}
\centering
\includegraphics[width=\columnwidth]{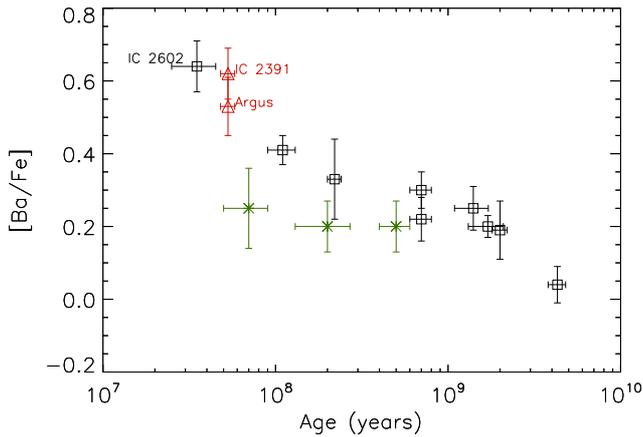}
\caption{[Ba/Fe] as a function of the open cluster's age for the sample from D'Orazi et al. (2009, empty squares); D'Orazi et al. (2012, starred symbols); and 
\citealt{desilva13} (triangles for IC 2391 and Argus).}\label{f:barium}
\end{figure}

\begin{figure*}
\centering
\includegraphics[width=\textwidth]{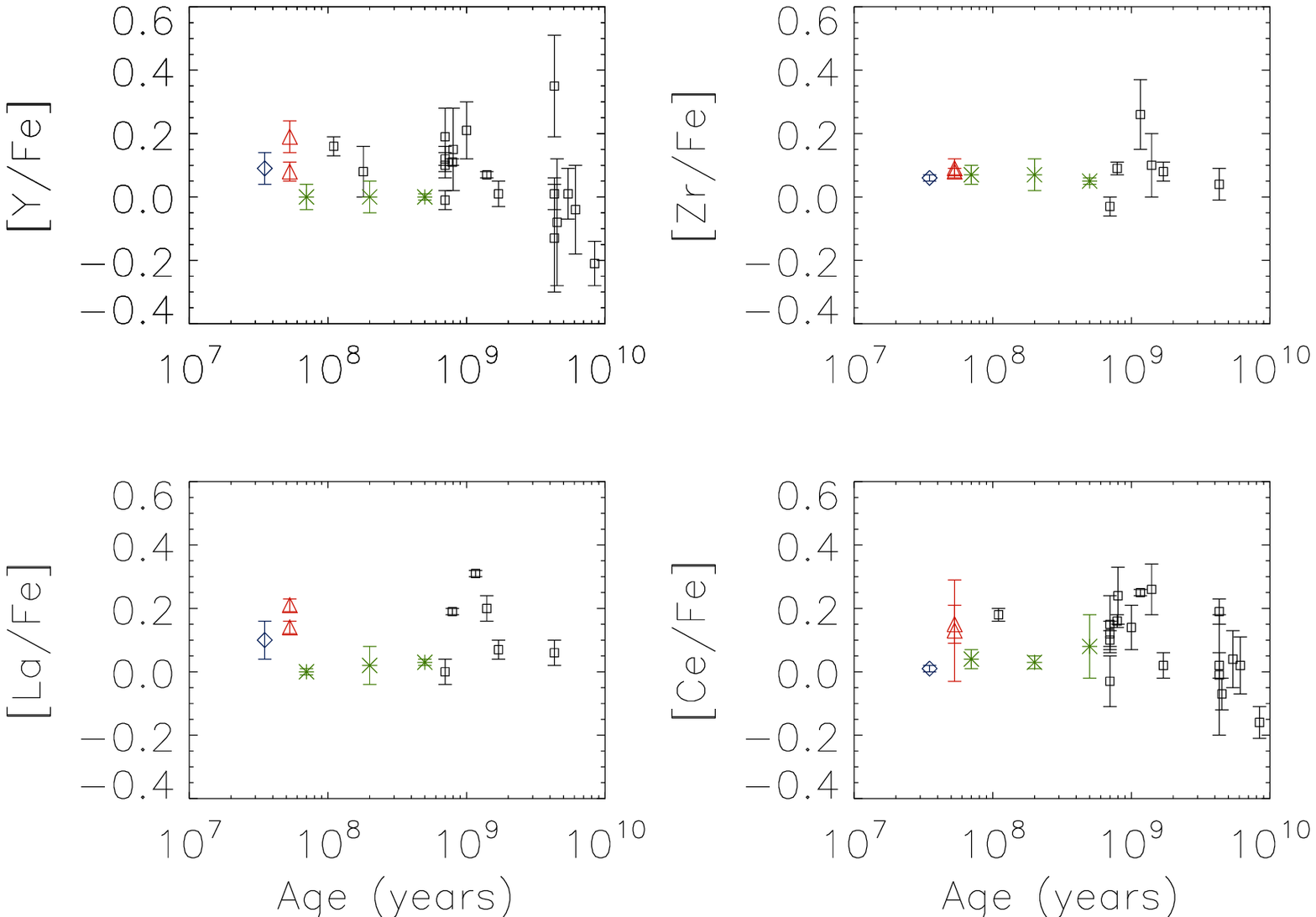}
\caption{Abundances of Y, Zr, La, and Ce  as a function of cluster age. Data points are from Maiorca et al. (2011, empty squares), D'Orazi et al. (2012, starred symbols), and the present study
(red empty triangles and blue diamond for IC2391/Argus and IC 2602, respectively).}\label{f:sproc}
\end{figure*}

The bizarre nature of these young stellar populations is  remarkable: {while Ba abundances are  enhanced by more than a factor of four above the solar value (see Table~\ref{t:results}), all the other heavy species exhibit solar scaled abundance patterns}. 
Thus, in agreement with what was found for slightly older associations (age $\sim$ 100 Myr) by \cite{dorazi12} and \cite{reddy15}, we  still face a barium conundrum. Crucially, the behaviour is much more extreme in this case, compared to the young local associations, because for those objects Ba abundances were at 0.2 - 0.3 dex level, whereas IC 2602, IC 2391 and Argus show values up to $\sim$ 0.65 dex. 
This can also be seen in Figure~\ref{f:barium}, where [Ba/Fe] is shown as a function of cluster age, including only measurements from our group, in order to minimise the scatter due to the combination of data from different sources. 
There is no doubt that [Ba/Fe] ratios are a sharp function of the age; however,  whether this corresponds to a real increase in the Ba abundance or to an artificial analysis effect is still a matter of debate. 
In our recent work (\citealt{dorazi12}) we have thoroughly investigated a variety of issues related to the presence of possible correlations of Ba with chromospheric effect, rotation, and departures from the LTE. However, we did not detect any significant correlation between Ba abundance and the chromospheric activity index logR$_{HK}$ or with X-ray excess related to the presence of hot coronae. Very recently we  published a comprehensive reassessment of the fundamental properties for the star GJ504. This star, which has been found to host a  substellar companion, presents a very curious case of discrepancy between the various age indicators. In particular, although it seems to be quite old according to spectroscopy (i.e. the gravity is lower than a main sequence star) and HR diagram, it  shows a significant level of chromospheric activity, compatible with an age of $\sim$ 200 Myr. Still, even though it exhibits this feature, we were able to  derive a solar Ba abundance; thus, our result seems to suggest that  the age causes high Ba abundances and not the chromospheric activity (see \citealt{dorazi16} for further details). 
Moreover, it is known that the Ba~{\sc ii} lines under investigation are  not  affected by significant non-LTE effects when considering the parameter space of our sample stars, i.e. solar-type stars (see also \citealt{korotin15}). \cite{reddy15} suggested that a viable alternative would be to exploit the Ba~{\sc i} line at 5535 \AA; however, this feature gives a solar Ba abundance for stars with effective temperatures hotter than about 5800 K, but increasingly subsolar Ba abundances (up to 0.5 dex) for cooler stars. 
As the authors have suggested, this is a strong indication of departures from LTE; thus, until corrections are provided for this line we use caution in  following this novel approach.
In addition, the authors also discuss the microturbulence effect (see also \citealt{dorazi12}). As we previously mentioned, Ba lines are near or on the flat part of the curve of growth, and hence very sensitive to microturbulence velocities. However, though playing with $\xi$ might  solve the Ba issue in part (a change in +0.5 km~s$^{-1}$ would imply a downward revision of 0.3 dex), it still would not solve the conundrum of these very young clusters where Ba is more than 0.65 dex in several cases.

From a nucleosynthesis model perspective, it is not possible to reconcile this behaviour with either the $s$- or the $r$-processes which have been suggested as being responsible for the production of elements heavier than iron. It is well established that the major contributors to the $s$-process are low-mass (M$\lesssim$ 4 M$_\odot$) asymptotic giant branch (AGB) 
stars, activating the $^{13}$C($\alpha$, $n$)$^{16}$O reaction to provide neutrons for the neutron captures on iron seeds (\citealt{busso99}; \citealt{kappeler11}). 
 We mention in passing that other sites of production,  e.g. the so-called $weak$ $s$-process component\footnote{The $weak$ $s$-process component likely comes from the He burning in the cores of massive stars (greater than approximately 15 M$_\odot$, e.g. \citealt{truran77}; \citealt{lamb77}) where the temperature is high enough for the $^{22}$Ne($\alpha$, $n$)$^{25}$Mg reaction to produce a substantial number of neutrons. These stars also have strong winds that eject this material into the interstellar medium.}, play a minor role in solar metallicity (see e.g. \citealt{travaglio04}).
None of the $s$-process models can explain a [Ba/La] larger than $\sim$ 0.15 dex (\citealt{mishenina15}).

Recently, another very interesting explanation has been suggested by \cite{mishenina15}. In order to reconcile the curious abundance pattern displaying significant Ba overabundances without similar enhancements in the other $s$-process species (especially La and Ce), they envisaged a different mechanism for the production of heavy elements, namely the intermediate $i$-process (\citealt{cowan77}; \citealt{herwig11}). This process might be at work in low-metallicity AGB stars (\citealt{hampel16}), post-AGB objects (\citealt{herwig11}; \citealt{lugaro15}), and super-AGB stars (\citealt{jones16}) and may produce [Ba/La] ratios up to $\approx$ 1.5 dex (see Figure 6 in \citealt{mishenina15} and \citealt{bertolli13}). 
Although a quantitative comparison with observations is still hampered by limitations related to uncertainties in theoretical models, a qualitative agreement suggests that our value of [Ba/La] $\sim $ 0.6 dex, though smaller than predictions, would be still in agreement when considering some dilution with material of $s$-process composition with [Ba/La]$\sim$ 0.
Recently, \cite{hampel16} has presented model predictions for the $i$-process in order to explain the heavy element pattern in the so-called carbon enhanced metal-poor (CEMP) $s$/$r$ stars; from their Figure 2 it is clear that Ba is  produced more than any other element, with the exception of the third-peak elements Pb and Bi. However, it is fundamental to stress that this Pb/Bi production strongly depends on the neutron exposure, which is substantially a free parameter in the models, so that the $i$-process can come in different ``flavours'' (see e.g. \citealt{roederer16} where instead a low neutron exposure has been adopted). Thus, a different source at different metallicity (CEMP stars are obviously much more metal-poor than our targets) might result in different yields.
We mention in this context that Pb abundances cannot be derived for our target stars because the blue spectral regions including the Pb~{\sc i} lines at 3683 and 4058 \AA~are very crowded and the spectral features very weak so even moderate rotational velocities cause severe spectral blending. In summary, this appealing explanation cannot be currently excluded and further theoretical investigations are needed.

\begin{sidewaystable*}
\centering
\caption{Stellar parameters, metallicity (from Fe~{\sc ii} lines), and barium abundances as reported by \cite{desilva13} and \cite{dr09} for our sample stars
belonging to the IC 2391/Argus associations and IC 2602, respectively. 
Abundances for Y, Zr, La, and Ce are from the present work.}\label{t:results}
\hspace*{-2.0cm}
\tabcolsep=0.17cm
\begin{tabular}{lcccccccccccccc}
\hline\hline
Star & $T_{\rm eff}$ & log$g$ & $\xi$ & [Fe/H] & [Ba/H] & [Ba/Fe] & [Y/H] & [Y/Fe] & [Zr/H] & [Zr/Fe] & [La/H] & [La/Fe] & [Ce/H] & [Ce/Fe]\\
 \hline 
        &                      &             &          &                            &            &     & & & & &      &           &           &             \\
        &                     &              &          &     &        &   &    &    {\bf IC 2391}              & &  &      &          & &   \\ 
        &                    &              &          &                             &            &                 & & & & &      &            &            &           \\
PMM 1142 &  5700 &  4.30 &  1.50 & $-$0.05$\pm$0.11 & 0.65$\pm$0.15 & 0.70$\pm$0.13  &  ~~0.04$\pm$0.13  & 0.09$\pm$0.09   & ~~0.00$\pm$0.12 &0.05$\pm$0.08   & 0.07$\pm$0.11 & 0.12$\pm$0.08 & ~~0.04$\pm$0.15 & 0.09$\pm$0.11 \\
PMM 4362  & 5650 &  4.30 &    1.40 & $-$0.08$\pm$0.09 & 0.52$\pm$0.20 & 0.60$\pm$0.17   &  $-$0.03$\pm$0.15 &  0.05$\pm$0.10       & ~~0.03$\pm$0.15 & 0.11$\pm$0.10   & 0.05$\pm$0.12 &  0.13$\pm$0.08 & ~~0.08$\pm$0.13 & 0.16$\pm$0.10\\
PMM   665   &  5500 &  4.50 &  1.60 & $-$0.11$\pm$0.12 & 0.35$\pm$0.15 & 0.46$\pm$0.12  &  $-$0.01$\pm$0.13  &   0.10$\pm$0.09     & $-$0.01$\pm$0.14  &0.10$\pm$0.10    &  0.05$\pm$0.12 & 0.16$\pm$0.08  & ~~0.09$\pm$0.13 & 0.20$\pm$0.10\\
      & & & & & & & & & & &  & & &  \\
      &         &            &         & & &                    &        &        {\bf ARGUS}   &     & & &    &   &   \\
      & & & & & & & & & \\
CD-283434 &  5600 & 4.30 & 1.50 & $-$0.10$\pm$0.08 & 0.40$\pm$0.15 &  0.50$\pm$0.12   & ~~0.04$\pm$0.12 & 0.14$\pm$0.09 &  $-$0.02$\pm$0.15  & 0.08$\pm$0.10     & 0.09$\pm$0.12 & 0.19$\pm$0.09 &  $-$0.09$\pm$0.12 & 0.01$\pm$0.09\\
CD-395833 & 5500 & 4.60  & 1.60 & $-$0.04$\pm$0.09 & 0.56$\pm$0.20 &  0.60$\pm$0.18   &  ~~0.17$\pm$0.16 & 0.21$\pm$0.13  & ~~0.03$\pm$0.15   &  0.07$\pm$0.10    & 0.18$\pm$0.20 & 0.22$\pm$0.18 & ~~0.20$\pm$0.20 & 0.24$\pm$0.18\\
     &         &             &                            &                     &     &      & & & & &           & & \\
       &         &            &                            &        &        & & & {\bf IC 2602}   &     &    &  & &  &   \\
      & & & & & & & & & \\
R66    &     5590 & 4.45 & 1.15 & $-$0.01 & 0.67$\pm$0.15 & 0.68$\pm$0.13 & ~~0.04$\pm$0.15 & 0.05$\pm$0.12  &  ~~0.05$\pm$0.12  & 0.06$\pm$0.08 & 0.13$\pm$0.12 & 0.14$\pm$0.08 & 0.00$\pm$0.13 & 0.01$\pm$0.09\\
R70 &       5760 & 4.45 & 1.10 & ~~0.00 & 0.68$\pm$0.15 &  0.68$\pm$0.13  & ~~0.12$\pm$0.13 & 0.12$\pm$0.10  &  ~~0.06$\pm$0.13 & 0.06$\pm$0.08 & 0.05$\pm$0.15 & 0.05$\pm$0.10  & 0.02$\pm$0.14 & 0.02$\pm$0.11\\
          & & & & & & & & & \\
 \hline\hline
\end{tabular}
\end{sidewaystable*}



\section{Concluding remarks}\label{sec:summary}

We have presented in this paper the determination of $s$-process elements Y, Zr, La, and Ce for three young, pre-main sequence clusters: IC 2391, the Argus association, and IC 2602. The aim of this work was to investigate the chemical pattern of the heavy species in very young stars for which barium abundances have been found to be enhanced up to $\sim $ 0.65 dex. We have exploited high-resolution, high S/N spectra and carried out a differential abundance analysis via spectral synthesis techniques, consistent with our previous works. 
Our results indicate that, at variance with what is expected according to the $s$-process nucleosynthesis, all the other $s$-process species do not follow Ba, but instead they show solar scaled ratios.
These targets hence confirm and extend to younger ages what was previously found for local associations (age $\sim$ 100 Myr) by \cite{dorazi12} and \cite{reddy15}, where only Ba exhibits enhanced values while all other $s$-process species are solar.
We have discussed several possible explanations for this curious behaviour, including chromospheric effects, microturbulence velocities,  departures from the LTE approximation, or the activation of a different nucleosynthesis source such as the $i$-process, as has  recently been suggested by \cite{mishenina15}. This last explanation seems to be very promising, but further theoretical investigations are needed, especially because of the uncertainties related to the models (see \citealt{mishenina15}; \citealt{hampel16}; \citealt{roederer16}). The barium puzzle is still not solved and remains perhaps the most intriguing chemical feature of young clusters.

\begin{acknowledgements}
This work made extensive use of the SIMBAD, WEBDA, and NASA ADS databases.
VD acknowledges support from the AAO distinguished visitor program 2016.
We thank Marco Pignatari and Maria Lugaro for very useful discussions.
We thank the anonymous referee for a very careful reading of the manuscript and for valuable comments and suggestions. 
\end{acknowledgements}

%
%

\bibliographystyle{aa} 


\end{document}